\PassOptionsToPackage{sort&compress}{natbib}
\documentclass[final,3p,times]{elsarticle}

\usepackage{amssymb}
\usepackage{amsmath}
\usepackage[breaklinks,colorlinks,citecolor=citcolor,urlcolor=blue,linkcolor=lcolor]{hyperref}

\newcommand{\Dd}{\mathrm{d}}
\newcommand{\bx}{\mathbf{x}}
\newcommand{\bs}{\mathbf{s}}
\newcommand{\bv}{\mathbf{v}}

\begin{document}

\begin{frontmatter}

\title{Generative diffusion models for lattice field theory}

\author[bielefeld]{Gert Aarts\footnote[1]{Based on the talk presented by GA at XQCD2026, Busan, Korea, 14-17 July 2026, \href{mailto:gert.aarts@uni-bielefeld.de}{gert.aarts@uni-bielefeld.de}}}
\author[swansea]{Diaa E.~Habibi}
\author[riken,utokyo]{Lingxiao Wang}
\author[CUHK]{Kai Zhou}

\affiliation[bielefeld]{organization={Faculty of Physics, Bielefeld University},
             city={Bielefeld},
             postcode={D-33615},
             country={Germany}}

\affiliation[swansea]{organization={Centre for Quantum Fields and Gravity, Department of Physics, Swansea University},
             city={Swansea},
             postcode={SA2 8PP},
             country={United Kingdom}}

\affiliation[riken]{organization={RIKEN Center for Interdisciplinary Theoretical and Mathematical Sciences (iTHEMS)},
             city={Wako},
             state={Saitama},
             postcode={351-0198},
             country={Japan}}

\affiliation[utokyo]{organization={Institute for Physics of Intelligence, Graduate School of Science, The University of Tokyo},
             addressline={Bunkyo-ku},
             city={Tokyo},
             postcode={113-0033},
             country={Japan}}

\affiliation[CUHK]{organization={School of Science and Engineering, The Chinese University of Hong Kong, Shenzhen (CUHK-
Shenzhen)},
             addressline={Guangdong},
             city={Shenzhen},
             postcode={518172},
             country={China}}

\begin{abstract}
Generative diffusion models offer a new route to ensemble generation in
lattice field theory. The backward, or denoising, process underlying these
models is a stochastic process with a learned, time-dependent drift, which
leads to a clear connection with stochastic quantisation. After reviewing
this correspondence, I turn to theories with a sign problem. There the complex
Boltzmann weight obstructs direct sampling, but complex Langevin dynamics can sometimes evade the sign problem by generating configurations on a complexified manifold, distributed according to a real and semi-positive distribution. This distribution is not known in closed form. I
discuss how this distribution can be learned from complex Langevin data using
score-based and, in particular, energy-based diffusion models. The latter give
direct access to the unnormalised distribution itself, which can subsequently
be used as the weight in a standard Markov chain Monte Carlo simulation.
Results are shown for a simple quartic model with a complex action.
\end{abstract}


\end{frontmatter}

\section{Introduction}
\label{sec:intro}

Machine learning methods and especially generative models are being heavily investigated as promising additions to the lattice toolbox; see
Refs.~\cite{Cranmer:2023xbe,Aarts:2025gyp,Tomiya:2025mlq,Kanwar:2026nlw} for reviews. The motivation is
familiar: critical slowing down and topological freezing
\cite{DelDebbio:2004xh,Schaefer:2010hu} limit what local updating algorithms
can achieve as the continuum limit is approached. Generative models have the potential to change the
paradigm \cite{Albergo:2019eim}. Instead of a single long Markov chain, with thermalisation followed
by a correlated evolution in equilibrium, one learns a map between a simple
prior and the complicated target distribution, and then generates many short,
independent trajectories. This is actively being pursued using, e.g., normalising flows
\cite{Albergo:2019eim,Kanwar:2020xzo} and stochastic normalising flows
\cite{Caselle:2022acb,Bonanno:2025pdp}.

Diffusion models
\cite{sohldickstein2015deepunsupervisedlearningusing,song2021scorebasedgenerativemodelingstochastic}
implement the same idea differently. Data are destroyed by a stochastic forward
process which adds noise, and new configurations are generated by a stochastic
backward process which removes the noise. The drift in the backward process is the
\emph{score}, the gradient of the logarithm of the time-dependent distribution, which is learned by a neural network. For a lattice field theorist, the
backward process is immediately recognisable as a Langevin equation, which
places diffusion models close to stochastic quantisation
\cite{Parisi:1980ys,Damgaard:1987rr}. This observation was made in
Ref.~\cite{Wang:2023exq}, where diffusion models were first applied in lattice
field theory. Below I review that correspondence (Sec.~\ref{sec:dmsq}) and then
turn to the main topic, theories with a sign problem
(Secs.~\ref{sec:complex}--\ref{sec:quartic}), following
Ref.~\cite{Aarts:2025kvh}.

\section{Diffusion models and stochastic quantisation}
\label{sec:dmsq}

Consider a theory with action $S(\phi)$ and target distribution
$p_0(\phi)\sim e^{-S(\phi)}$, from which a training ensemble of configurations can be sampled. In the forward process the fields evolve
according to
\begin{equation}
  \frac{\partial\phi(x,t)}{\partial t} = \tfrac{1}{2}K[\phi(x,t),t] + g(t)\eta(x,t),
  \qquad 0\leq t\leq T,
\label{eq:forward}
\end{equation}
with $\phi(0)$ drawn from the training ensemble, $t$ the diffusion time, and $\eta$ Gaussian noise. The drift $K$
may be set to zero; the noise strength $g(t)$ is chosen such that at $t=T$ the
distribution $p_T$ is a simple, structureless prior (see, however, Ref.~\cite{Aarts:2024rsl}). New configurations are
generated by running the backward process from $t=T$ down to $t=0$,
\begin{equation}
  \frac{\partial\phi(x,t)}{\partial t} = \tfrac{1}{2}K[\phi(x,t),t] - g^2(t)\nabla_\phi\log p(\phi,t)
                + g(t)\eta(x,t),
\label{eq:backward}
\end{equation}
in which the score $\nabla_\phi\log p(\phi,t)$ appears as a time-dependent
drift, guiding the trajectories back to the target distribution.  
This score is not known and is approximated by a neural network,
$\bs_\theta(\phi,t)\approx\nabla_\phi\log p(\phi,t)$, obtained by minimising
\begin{equation}
  \mathcal{L}(\theta,\lambda) = \frac{1}{2}\int_0^T\!\Dd t\;
    \mathbb{E}_{p(\phi,t)}\Big[\lambda(t)\,
    \big\|\bs_\theta(\phi,t)-\nabla_\phi\log p(\phi,t|\phi_0)\big\|^2\Big],
\label{eq:loss}
\end{equation}
where the intractable score has been traded for the conditional score of the
forward kernel \cite{6795935} and $\lambda(t)$ is a weight function.

The connection with stochastic quantisation is now immediate. Writing
$p(\phi,t)=e^{-S(\phi,t)}/Z$, so that
$\nabla_\phi\log p = -\nabla_\phi S$, and introducing $\tau=T-t$ as the time for the backward evolution, the generative process
(\ref{eq:backward}) with $K=0$ reads
\begin{equation}
  \frac{\partial\phi(x,\tau)}{\partial\tau} =
    -g^2(T-\tau)\nabla_\phi S(\phi,T-\tau) + g(T-\tau)\eta(x,\tau),
\label{eq:backwardS}
\end{equation}
to be compared with the Langevin equation of stochastic quantisation,
\begin{equation}
  \frac{\partial\phi(x,\tau)}{\partial\tau} =
    -\nabla_\phi S(\phi) + \sqrt{2}\,\eta(x,\tau),
\label{eq:sq}
\end{equation}
whose associated Fokker--Planck equation has the stationary solution
$p(\phi)\sim e^{-S(\phi)}$. The structure is the same; the difference lies in
how the equations are used \cite{Wang:2023exq}. In stochastic quantisation the
drift is fixed and determined by the known action, the noise variance is
constant (though it may be generalised using kernels), and one thermalises and
then evolves for a long time in equilibrium. In a diffusion model both drift
and noise variance are time dependent, the drift is \emph{learned from data}
rather than derived from an action, and the evolution takes place over the
finite interval $0\leq\tau\leq T$, with many short runs replacing one long one.
A diffusion model is thus an alternative route to generate the same ensembles, and the
experience accumulated in stochastic quantisation -- kernels, discretisation
effects, higher-order integrators, the Fokker--Planck analysis -- can be
immediately applied \cite{Wang:2023exq,Fukushima:2024oij,Hirono:2024zyg}.

\begin{figure}[t]
\centering
 \includegraphics[width=0.5\linewidth]{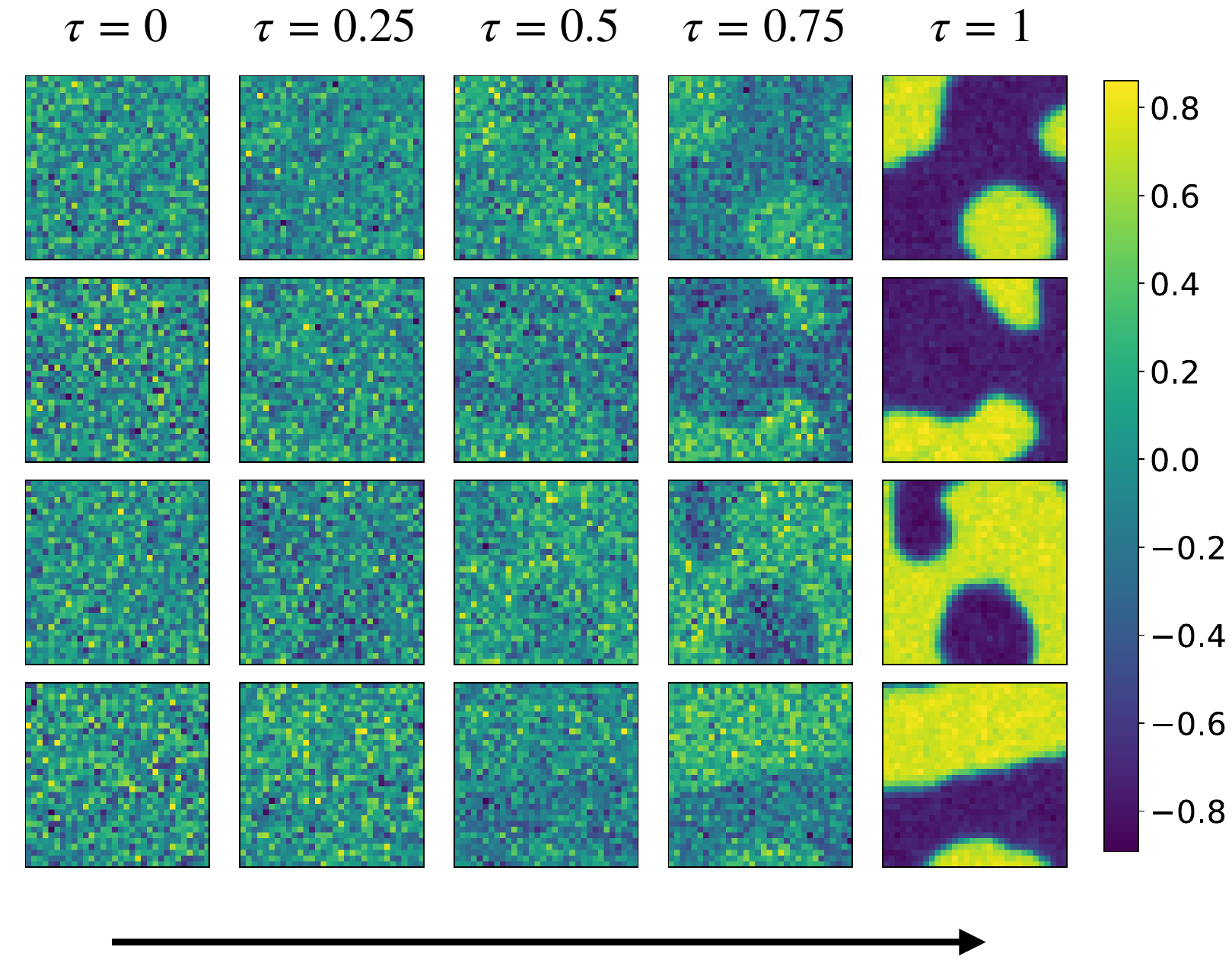}
\caption{Denoising (backward) process for the two-dimensional $\phi^4$ theory
in the broken phase on a $32^2$ lattice, at $\tau=0$ (the prior), $0.25,0.5,0.75$ and $1$ (the target). From Ref.~\cite{Wang:2023exq}.}
\label{fig:phi4}
\end{figure}

The first application in lattice field theory \cite{Wang:2023exq} was to the
two-dimensional scalar $\phi^4$ theory on a $32^2$ lattice, with training data
generated by Hybrid Monte Carlo (HMC) in the symmetric and broken phases. The
backward process starts from noise and, in the broken phase, large-scale
clusters emerge, as expected on physical
grounds (Fig.~\ref{fig:phi4}).

Since then the approach has been extended to abelian 
 \cite{Zhu:2025pmw,Vega:2025hgz} and non-abelian gauge theories
\cite{Aarts:2026zzr,Vega:2025xma,Alharazin:2026lcb}.
In Ref.~\cite{Aarts:2026zzr}, discussed at this meeting by Thomas Ranner, a
gauge-equivariant diffusion model for two-dimensional U(2) and SU(2) gauge
theory is built from lattice gauge equivariant convolutional neural networks
\cite{Favoni:2020reg} and sampled with the Metropolis-adjusted annealed
Langevin algorithm \cite{Zhu:2025pmw}.
Trained on a \emph{single} ensemble, it generalises to
larger inverse coupling and larger volumes at moderate acceptance rates, with
Wilson loops and the topological susceptibility reproduced with little loss of
accuracy. In fact, there has been considerable activity and I refer to Ref.~\cite{Aarts:2026yim} for an extensive list of references as well as a detailed discussion of diffusion models for non-abelian theories.

\section{Theories with a complex action}
\label{sec:complex}

I now turn to theories in which the weight
$\rho(x) = e^{-S(x)}/Z$, with $Z=\int\Dd x\, e^{-S(x)}$, is complex, so that
$\rho$ is not a probability distribution and importance sampling is
unavailable (instead of a field theory, I consider one degree of freedom, $x$, only). This is the sign problem; see Ref.~\cite{Aarts:2026ssp} for a
recent overview of the methods devised to deal with it. 
One such method is
complex Langevin (CL) dynamics, in which the real Langevin process
(\ref{eq:sq}) is continued into the complexified configuration space,
$x\to z=x+iy$,
\begin{equation}
  \dot z(t) = K[z(t)] + \sqrt{2}\,\eta(t),
  \qquad
  K(z) = \frac{\partial}{\partial z}\log\rho(z) = -\frac{\partial S(z)}{\partial z}.
\label{eq:cl}
\end{equation}
In terms of real variables, this is a two-dimensional real process,
\begin{equation}
  \dot x = K_x(x,y) + \sqrt{2N_x}\,\eta_x, 
  \qquad
  \dot y = K_y(x,y) + \sqrt{2N_y}\,\eta_y,
  \qquad
  K_x=\mathrm{Re}\,K(z)\big|_{z\to x+iy}, 
  \qquad
  K_y=\mathrm{Im}\,K(z)\big|_{z\to x+iy}.
\label{eq:clxy}
\end{equation}
The noise satisfies $N_x-N_y=1$; in practice real noise,
$N_x=1$, $N_y=0$, is used \cite{Aarts:2009uq}. The associated Fokker--Planck equation,
\begin{equation}
  \partial_t\, p(x,y;t) =
    \big[\partial_x(N_x\partial_x - K_x) +
         \partial_y(N_y\partial_y - K_y)\big]\, p(x,y;t),
\label{eq:fp2d}
\end{equation}
is \emph{real}, and so is its solution $p(x,y;t)$, which is semi-positive and
may be read as a probability distribution on the complexified manifold.
Expectation values follow as
$\langle O[x(t)+iy(t)]\rangle_\eta = \int\Dd x\,\Dd y\; p(x,y;t)\,O(x+iy)$,
and the process yields the correct answer provided
\begin{equation}
  \int \Dd x\,\Dd y\; p(x,y)\, O(x+iy) = \int \Dd x\; \rho(x)\, O(x)
\label{eq:correctness}
\end{equation}
for a sufficiently large set of observables; formally this is solved by
$\rho(x)=\int\Dd y\, p(x-iy,y)$. A detailed discussion on the validity of this approach was provided by Michael Mandl at this meeting 
\cite{Mandl:2026ebc}.

The real distribution $p(x,y)$ is therefore the central object: if it were
known, the sign problem would be solved as one could directly sample from it. This is
the aim of the wider programme of positive representations of complex measures
\cite{Weingarten:2002xm,Seiler:2017vwj}, which goes beyond complex Langevin
dynamics. In practice, however, $p(x,y)$ is not known, but can only be reconstructed as a histogram of
CL data. The question posed in Ref.~\cite{Aarts:2025kvh} is whether a diffusion
model can learn it, since in particular a diffusion model does not care where its training data come from, and CL trajectories provide samples from $p(x,y)$ just as HMC provides samples from $e^{-S}$.

\section{Score-based versus energy-based models}
\label{sec:ebm}

What the network in a diffusion model learns is the score $\bs_\theta(\bx,t)$, with $\bx=(x,y)$, and
not the distribution. For sampling this suffices, since only the score enters
the backward process (\ref{eq:backward}). To recover $p(\bx)$ one would like to
integrate the score, but in a score-based model (SBM) nothing enforces that
$\bs_\theta$ is a gradient. Using a Helmholtz decomposition,
$\bs_\theta = \nabla\Phi_\theta + \mathbf{r}_\theta$ with
$\nabla\cdot\mathbf{r}_\theta=0$, the squared norm in the loss (\ref{eq:loss})
splits as
\begin{equation}
  \big\|\bs_\theta - \nabla\log p\big\|^2 =
  \big\|\nabla\Phi_\theta - \nabla\log p\big\|^2 +
  \big\|\mathbf{r}_\theta\big\|^2 ,
\label{eq:split}
\end{equation}
so that the conservative and non-conservative parts are minimised independently. The
latter should vanish but at finite training does not, and line
integrals of $\bs_\theta$ are path dependent. In
two dimensions the obstruction is the curl of the learned score, which can be
monitored directly and is found to be small but non-zero \cite{Aarts:2025kvh}.

An energy-based model (EBM) avoids this by construction. Here the network
parametrises an energy,
\begin{equation}
  E_\theta(\bx,t) = \tfrac{1}{2}\big\|\bv_\theta(\bx,t)\big\|^2
    \approx -\log p(\bx,t),
\label{eq:ebm}
\end{equation}
which is manifestly semi-positive, and the (now conservative) score follows by differentiation,
\begin{equation}
  -\partial_i E_\theta(\bx,t) =
   -\bv_\theta(\bx,t)\cdot\partial_i \bv_\theta(\bx,t)
   \approx \partial_i\log p(\bx,t).
\label{eq:ebmscore}
\end{equation}
Training is more expensive, since gradients of the network enter the loss, and smooth activation functions (SiLU,
Mish) are required rather than the usual (leaky) ReLU. The reward is direct
access to the distribution itself, up to an overall normalisation,
$p_\theta(\bx)\sim\exp[-E_\theta(\bx)]$, with
$E_\theta(\bx)=E_\theta(\bx,t\to 0)$.

\section{Complex quartic model}
\label{sec:quartic}

\begin{figure}[t]
\centering
\includegraphics[width=0.44\linewidth]{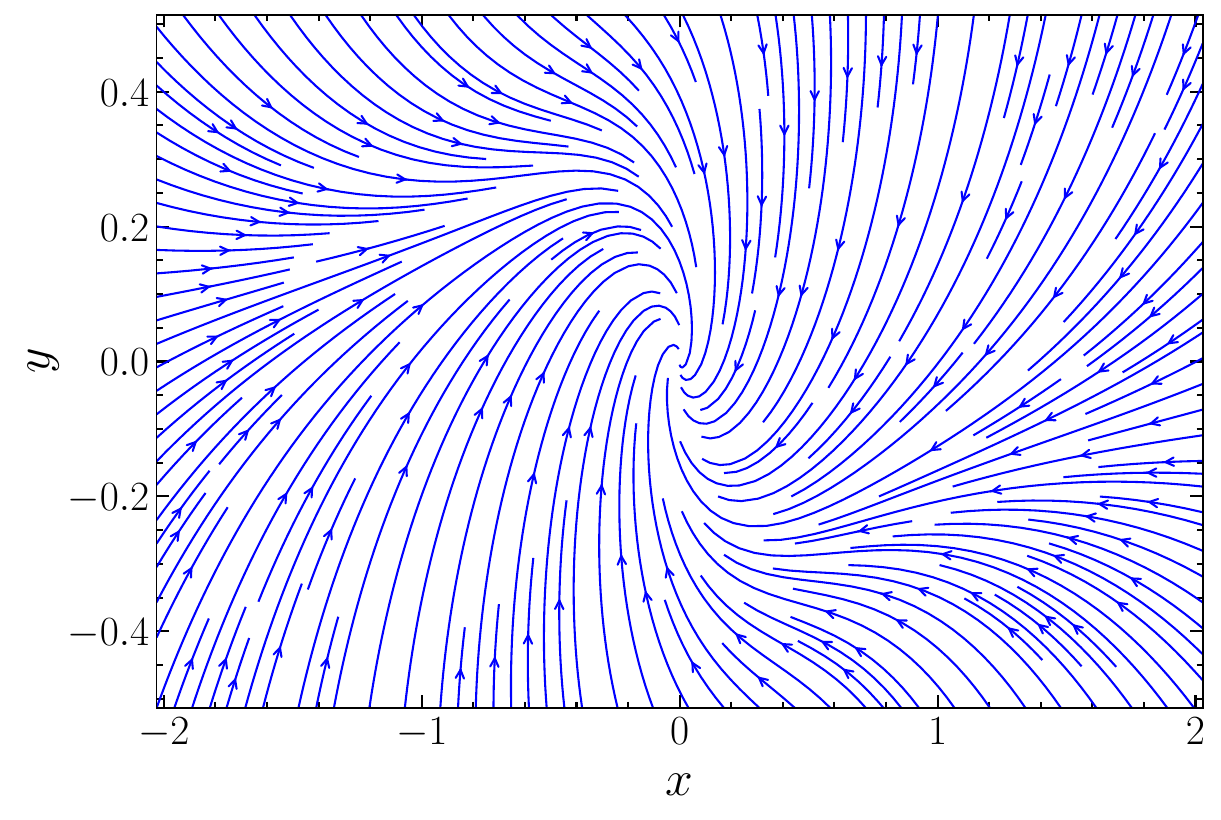} \;\;\;
\includegraphics[width=0.5\linewidth]{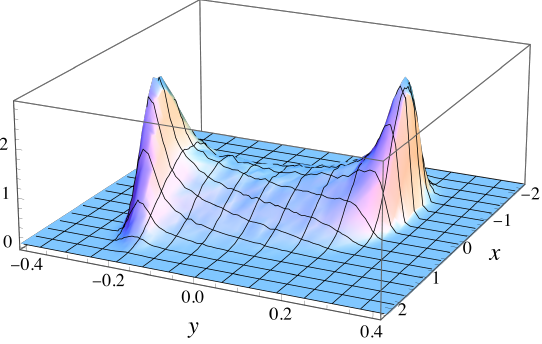}
\caption{Complex quartic model with $\sigma_0=1+i$, $\lambda=1$. Left: CL drift
in the complex plane. Right: histogram $p(x,y)$ obtained by binning CL data.
From Refs.~\cite{Aarts:2013uxa,Aarts:2025kvh}.}
\label{fig:cl}
\end{figure}

As a testing ground we take the quartic model with a complex coupling,
$S = \tfrac{1}{2}\sigma_0 x^2 + \tfrac{1}{4}\lambda x^4$, with $\sigma_0 = A + iB$,
a standard benchmark for complex Langevin dynamics \cite{Aarts:2013uxa}. The CL process converges to the
correct result provided $3A^2-B^2\geq 0$; the distribution $p(x,y)$ then has
compact support in the imaginary direction, confined to a strip $|y|<y_-$ with
$y_-^2 = (A/2\lambda)(1-\sqrt{1-B^2/3A^2})$. All results below use
$A=B=\lambda=1$, for which $y_-\simeq 0.3029$. The CL drift and the resulting
histogram are shown in Fig.~\ref{fig:cl}: the distribution is localised, with
two pronounced peaks, it satisfies $p(-x,-y)=p(x,y)$, and is strictly zero when $|y|>y_-$.

\begin{figure}[t]
\centering
\includegraphics[width=0.44\linewidth]{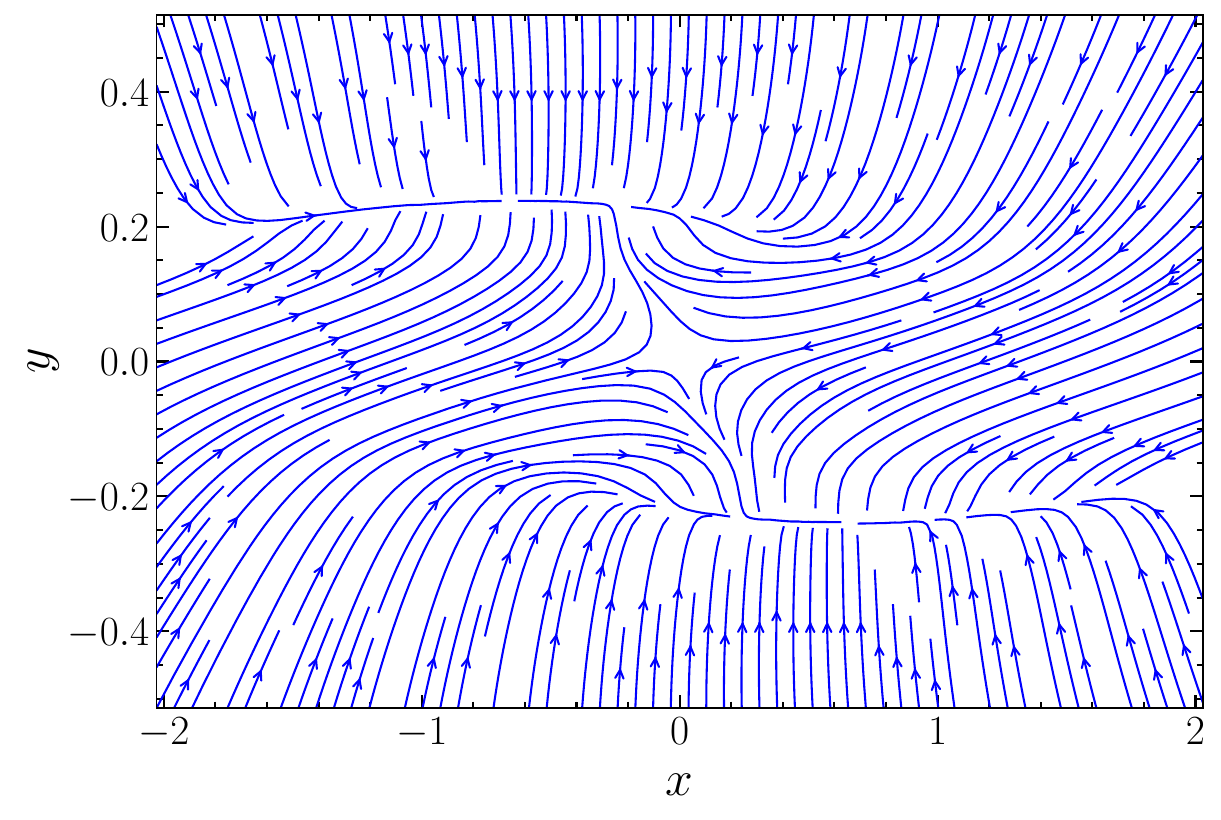} \;\;\;
\includegraphics[width=0.5\linewidth]{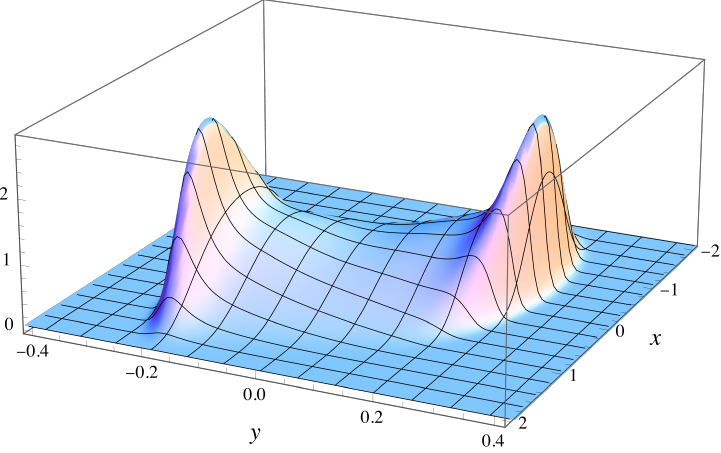}
\caption{Energy-based diffusion model trained on CL data. Left: minus the learned score
$\nabla E_\theta(\bx)$. Right: the corresponding distribution
$p_\theta(\bx)\sim\exp[-E_\theta(\bx)]$, to be compared with the right panel of
Fig.~\ref{fig:cl}. From Ref.~\cite{Aarts:2025kvh}.}
\label{fig:ebm}
\end{figure}

The CL data are used to train the diffusion models. Fig.~\ref{fig:ebm} shows
the score learned by the energy-based model and the corresponding distribution
$p_\theta(\bx)$. Comparing with Fig.~\ref{fig:cl} (right), the
two-peak structure, the width in the $y$ direction and the overall shape are
reproduced, the learned distribution being a smooth function rather than binned
data. The energy itself (not shown) has a simple and appealing structure: two degenerate,
shallow minima -- corresponding to the two peaks -- surrounded by steeply
rising barriers which enforce the localisation in the complexified space.

The final step follows from the observation that $E_\theta$ is a real function
encoded in a neural network, which can be evaluated anywhere within the domain
covered by the data. It can therefore serve as the weight in an ordinary
Metropolis simulation, with accept/reject step based on
$p_\theta(\bx)\sim\exp[-E_\theta(\bx)]$. Such a step is impossible for the
original complex weight $\rho(x)$; it is made possible here precisely because
the diffusion model has delivered a real and semi-positive distribution. 
Note that after training, the procedure makes no further reference to the CL process.

\begin{figure}[t]
\centering
\includegraphics[width=0.65\linewidth]{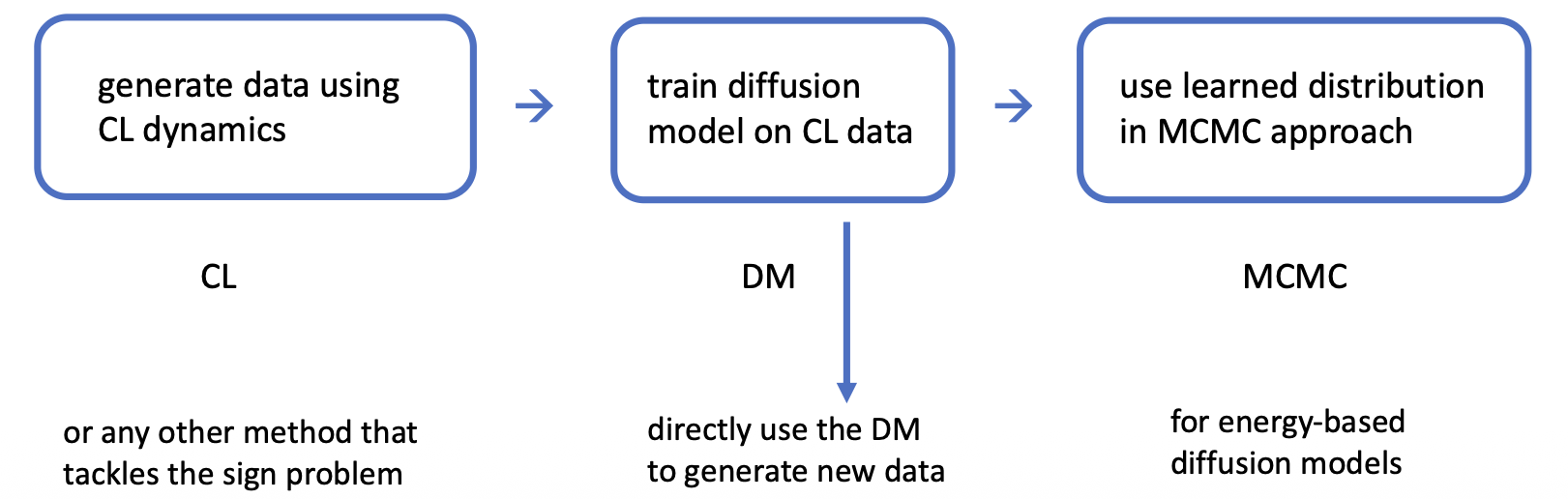}
\caption{Three-stage approach: configurations in the complexified space are
generated with complex Langevin dynamics, a diffusion model is trained on them,
and the learned energy is used as the weight in a standard Markov chain Monte
Carlo simulation.}
\label{fig:threestage}
\end{figure}

The resulting strategy is summarised in Fig.~\ref{fig:threestage} and a detailed comparison between the various approaches can be found in Ref.~\cite{Aarts:2025kvh}.
It should be stressed what this does and does not achieve. The diffusion model does not
repair a CL simulation which converges to the wrong result: it learns the
distribution the process actually samples, whatever that is. What it provides
is an explicit, differentiable representation of a distribution previously
available only as binned data, and with it the possibility of sampling in the
complexified space without referring to the complex weight again. The first
stage, moreover, is not tied to complex Langevin dynamics: any method which
produces configurations in a complexified or deformed space could serve as the
data source.

\section{Summary and outlook}
\label{sec:summary}

Generative diffusion models provide a new approach to ensemble generation in
lattice field theory, and since the approach is closely related to stochastic quantisation, it fits 
naturally within the existing theoretical framework. While high-quality ensembles are required for training, 
the generalisation seen recently \cite{Aarts:2026zzr,Aarts:2026yim} is encouraging for the future.

For theories with a sign problem, diffusion models offer a new perspective on
an old question: what is the real, semi-positive distribution sampled by
complex Langevin dynamics on the complexified manifold? Energy-based models
answer this constructively, delivering $-\log p(\bx)$ as a neural network which
can be evaluated anywhere and used as a Boltzmann weight in a conventional
Monte Carlo simulation. The natural next steps are to move from the
single-variable model studied here to field theories, where the dimensionality
of the complexified configuration space grows with the volume. There are many directions left to explore.

\section*{Acknowledgements}

\noindent
GA thanks the organisers of XQCD2026 for a stimulating meeting. GA also thanks his co-authors, 
Diaa Habibi, Lingxiao Wang and Kai Zhou, 
Andreas Ipp, David M\"uller and Thomas Ranner, and Wei Wang and Qianteng Zhu for the exciting collaboration on diffusion models.
GA was supported by STFC grant ST/X000648/1 and by a Royal Society Leverhulme Trust Senior Research Fellowship, and is currently supported by the W\"ubben Stiftung Wissenschaft.  
The first draft of this contribution was composed by Claude Opus 5 and subsequently edited by GA.


\setlength{\bibsep}{0.1ex}
\def\bibfont{\footnotesize}


\end{document}